
\magnification=\magstep1
 \baselineskip=12.67pt plus 2pt minus 1 pt 
 \hsize=6truein
 \hoffset=18pt
 \parskip=1pt plus 1pt
 \vsize=8truein
\count100=1
 \def\nullx{\hfill}
 \pageno=\count100
 \headline{\ifodd\pageno\rightheadline \else\leftheadline\fi}
 \footline{\ifnum\pageno=\count100{\hss}\else\nullx\fi}
 \def\rightheadline{\ifnum\pageno=\count100 \nullx%
\else\it\chptitle\hfil\rm\folio\fi}
 \def\leftheadline{\ifnum\pageno=\count100 \nullx%
\else\rm\folio\hfil\it\author\fi}
 \hfuzz=1pt
 \tolerance=10000

 
 \font\title=cmbx12

 \font\sc=cmcsc10
 \newif\ifninebanner\ninebannertrue 
 \newdimen\bannerskip
 \ifninebanner
 \bannerskip=8.5pt
 \font\brm=cmr9 scaled 833
 
 \font\bsl=cmsl9 scaled 833

 \else	
 \bannerskip=7.5pt
 \font\brm=cmr8 scaled 833
 
 \font\bsl=cmsl8 scaled 833

 \fi

\def\eth{{\mathsurround=0pt $\delta$\kern-0.6em\raise0.6ex\hbox{$-$}}}
\def\thorn{{\mathsurround=0pt $\flat$\kern-0.44em\lower0.17ex\hbox{$\vert\;$}}}
\def\beginflushfootnote{\bgroup\parindent=0pt\footnote}
\def\endflushfootnote{\egroup}
\def\n{\noindent}
\hfuzz=1pt
\def\chptitle{}
\def\author{}
\null\vskip72pt
\def\frac#1#2{{#1\over #2}}
\def\e{{\rm e}}

\rightline{hep-th/9508024}
\rightline{ZU-TH 1/95}
\vskip10pt
\centerline{\title Gravitating, Gravitational and Dilatonic  Sphalerons}
\vskip12pt
\centerline{\sc George Lavrelashvili}\smallskip

\centerline{\sl Institut f\"ur Theoretische Physik}
\centerline{\sl Universit\"at Z\"urich}
\centerline{\sl Winterthurerstrasse 190}
\centerline{\sl CH-8057 Z\"urich, Switzerland }

\beginflushfootnote{}{{\baselineskip\bannerskip\brm
\noindent Talk given at the
{\bsl Heat Kernel Techniques and Quantum Gravity},
Winnipeg, Canada, August 2-6, 1994.
To be published in the Proceedings.
\par}}\endflushfootnote
\vskip24pt

{\narrower\medskip \noindent {\bf Abstract.} In the present contribution we
shall give a brief review of the main properties of sphalerons
in various theories with a Yang--Mills field.
\bigskip}

\n {\bf 1. Introduction}

As is well known, the
vacuum in gauge theories has a complicated structure
[1].  One can label vacua with a topological (Chern--Simons)
number.
Vacua with different topological numbers are
separated by a potential barrier, whose
height is set by the sphaleron [2], [3].

Topologically nontrivial fluctuations
of the gauge field lead to fermion number nonconservation [4]
via the anomaly [5]. The fermion number nonconservation in
topologically nontrivial gauge field backgrounds can be described by the
generalized Bogolyubov transformation technique [6], [7]
or in terms of a level-crossing picture [8].
The rate of the  fermion number nonconservation
depends on the energy of the process.
The sphaleron solution determines the energy scale for processes
with a strong nonconservation of fermion number.

Recently  a discrete series of static, spherically symmetrical
solutions of Einstein--Yang--Mills (EYM) theory was found
was found by Bartnik and McKin-\break non [9].
It turns out that Bartnik--McKinnon (BMK)
solutions are gravitational analogues
of the electroweak sphaleron [10], [11].
Solutions of the same nature were found in Yang--Mills-dilaton
(YMD) theory [12], [13].
It was understood that the existence of all of these solutions
is related to topological properties of the YM field
configuration space.

In the present contribution we give a brief
review of different YM sphalerons.

In the next section we describe sphaleron solutions in
YMH, EYM and YMD theories.
In section 3 we discuss the main properties
of these solutions and their interpretation.
In section 4 we discuss similarity and difference
in various sphalerons.
Section 5 contains concluding remarks.\bigskip

\n {\bf 2. Sphaleron solutions in different theories}

In the present section we shall discuss several
theories with the YM field in which sphaleron solutions exist,
namely YMH with the Higgs doublet, EYM, and YMD,
and describe the main properties of sphalerons.\medskip

\n {\bf 2.1. Electroweak sphaleron}

The electroweak sphaleron has a relatively long history.
It was found by
R.F. Dashen, B. Hasslacher, and A. Neveu (DHN) [14]
in 1974 in relation with hadron physics and
rediscovered later in the context of  nuclear physics
by  Boguta [15].
In 1983 an analysis of the properties of the configuration
space of an $SU(2)$ YM field [2] led to the claim of
existence of  saddle point solutions in YMH theory.
It was realized  [3]
that they coincide with the DHN solutions.

The argument for the existence of sphalerons runs as follows [2].
Let us consider a one-parameter family of configurations
interpolating between the vacua with  different topological
(Chern--Simons) numbers. The asymptotic behaviour of the gauge field
defines a map $S^2 \to SU(2)\simeq S^3$.
A suitable one-parameter family of maps is topologically
equivalent to a single nontrivial map $S^3 \to S^3$.
Denoting the maximum of the (static) energy along each path $l$
by $E(l)$, we may take the minimum of $E(l)$ running through all
nontrivial paths. This minimum corresponds to a saddle point
of the energy functional.

 In pure YM theory there is no scale and there are no static
solutions [16] in $(1+3)$ dimensions.
One way out is to introduce a Higgs field.

The action for the YMH theory has the form
$$ S_{\rm YMH}=
\int \left( -\, {1\over 4g^2} F^a_{\mu\nu}F^{a\ \mu\nu}
+{(D^\mu\Phi)}^{\dagger} D_\mu\Phi-
\lambda ({\Phi}^{\dagger}\Phi-{v^2\over 2})^2
\right)d^4x \eqno (1)$$
where $F^a_{\mu\nu}$ is the $SU(2)$ gauge field strength,
$F^a_{\mu\nu}=\partial_{\mu} W^a_{\nu}-
\partial_{\nu} W^a_{\mu}+\epsilon^{abc} W^b_{\mu}W^c_{\nu}$,
and $a=1,2,3$ is the $SU(2)$ group index, $\mu,\nu = 0,1,2,3$
are space-time indices.
Covariant derivatives are defined by
$D_\mu\Phi=\partial_\mu\Phi-{i\over 2}\tau^a W^a_\mu\Phi$.

We are interested in spherically symmetric solutions.
The most general spherically symmetric
ansatz for the {\sl SU(2)\/} Yang--Mills field $W_\mu^a$ can be written
(in the Abelian gauge) as [17]
$$\eqalignno{W_t^a &=(0,0,A_0)\,,\qquad  W_\theta^a
=(\phi_1,\phi_2,0)\,,\cr  W_r^a&=(0,0,A_1)\,, \qquad
 W_\varphi^a =(-\phi_2 \sin\theta,\phi_1\sin\theta,\cos\theta)\,.
 &(2)}$$
For the Higgs field we take
$$
\Phi ={v\over \sqrt{2}}\; [H+iK(\vec{n}\cdot\vec{\tau})]
{0\choose 1},\, \eqno(3)$$
where $\vec{\tau}$ are the usual Pauli isospin matrices
and $\vec{n}=\vec{x}/r$.

The ansatz (2) is form-invariant
under gauge transformations around the third isoaxis,
with $A_\alpha$ transforming as a $U(1)$ gauge field
on the reduced space-time $(t, r)$, whereas
$\phi=\phi_1+i\phi_2$ is a scalar field of charge one
with respect to the $U(1)$.
Introducing $\chi=H+iK$ we find
$$A_{\alpha}\to A_{\alpha}+\partial_{\alpha}\Omega,\qquad
\phi \to \e^{i\Omega}\phi, \qquad
\chi\to \e^{i\Omega/2}\chi.\eqno (4)$$
With respect to this $U(1)$ one may define the `charge conjugation'
$$A_\alpha\to -A_\alpha, \qquad
\phi\to\overline\phi, \qquad
\chi\to \overline{\chi}.\eqno (5)$$
The even sector with respect to this  charge conjugation is given by
$$A_0 =0,\; A_1 =0,\; \phi_1 \equiv W(r),\; \phi_2 =0, K=0 .\eqno (6)$$
In this sector the ansatz (2) is equivalent
to the usual ``monopole"  ansatz
$$W_0^a=0, \quad
W_i^a=\epsilon_{aij}{n_j\over r}(1-W(r))\eqno (7)$$
with  $n_j=x_j/r$.
And the Higgs field is
$$\Phi ={v\over \sqrt{2}}\; H(r)
{0\choose 1}\, .\eqno (8)$$
The reduced action in this sector has the form
$$\eqalignno{
S^{\rm red}_{\rm YMH}&= -\,
\frac{4\pi v}{g}\int\left(
\left({dW\over d\xi}\right)^2 +{(W^2-1)^2 \over 2\xi^2}\right. \cr
&\quad \left. {}+ {\xi^2\over 2} \left({dH \over d\xi}\right)^2
+{H^2 (1+W)^2\over 4}
+{1\over 4}{\lambda \over g^2} \xi^2 (H^2-1)^2
\right)d\xi&(9)}$$
where $\xi=gvr$.

The corresponding equations of motion are
$$\eqalignno{
{d^2 W\over d\xi^2}&= {W(W^2-1)\over {\xi}^2}+{H^2\over 4}
(1+W)\;, \cr
{d\over d\xi}\left({\xi}^2 {d H \over d\xi}\right) &=
{H (1+W)^2\over 2} +{\lambda\over g^2} {\xi}^2 (H^2-1) H \;.&(10)}$$
The solution has to interpolate between
$$ W=1 \;,\qquad  H=0\;,\eqno (11)$$
at $\xi\to 0$  and
$$ W=-1\;,\qquad H=1 \;\eqno (12)$$
for $\xi\to \infty$.

It was found [14], [3] that equations (10)
indeed	have a sphaleron solution $\{ W(r), H(r) \}$
which satisfies boundary conditions
(11),~(12).\medskip

\n {\bf 2.2. Gravitational sphaleron}

In 1988 Bartnik and McKinnon unexpectedly found
a discrete sequence of globally regular solutions
of the EYM theory.

We say unexpectedly, because neither vacuum Einstein nor
pure YM theory has  nontrivial globally regular, static, finite
energy solutions [16]. There are also no such solutions
in the EYM theory in $(2+1)$ dimensions [18].

The action for the EYM theory has the form
$$S_{\rm EYM}=
\frac{1}{4\pi}\int\left(-\frac{1}{4 G}R
-\, {1\over 4g^2} F^a_{\mu\nu}F^{a\ \mu\nu}
\right)\sqrt{-g}\, d^4x \eqno (13)$$
where \ $g$ denotes the gauge coupling constant
and $G$ is Newton's constant.

A convenient parametrization for the metric turns out to be
$$ds^2=S^2(r)N(r)dt^2-{dr^2\over N(r)}
  -r^2d\Omega^2\;, \eqno (14)$$
where $d\Omega^2=d\theta^2+\sin^2(\theta)d\varphi^2$
is the line element of the unit sphere.

For the $SU(2)$ YM potential we make the usual (`magnetic')
spherically symmetric ansatz (7).
Substituting this ansatz into the action
we obtain the reduced action
$$S^{\rm red}_{\rm EYM}=-\int S\left[{1\over 2G}(N+rN'-1)
+{1\over g^2}\left(N W'^2+{(1-W^2)^2\over2r^2}\right)\right]\,dr\;,
\eqno (15)$$
where a prime denotes $d\over dr$.

The resulting field equations are
$$\eqalignno{
 (NS W')'&= S
   {W(W^2-1)\over r^2} \;, \cr
 N' &= {1\over r}\left(1-N
 -2\left(N W'^2+{(1-W^2)^2\over2r^2}\right)\right)\;, \cr
 S^{-1}S'&= \frac{2W'^2}{r} \;.&(16)}$$
The field equations (16)  have
singular points at $r=0$ and $r=\infty$ as well
as  points where $N(r)$ vanishes.
Regularity at $r=0$ of a configuration
requires $N(r)=1+O(r^2)$, $W(r)=\pm1+O(r^2)$
 and $S(r)=S(0)+O(r^2)$.
Since $W$ and $-W$ are gauge equivalent we may choose
$W(0)=1$. Similarly we can assume $S(0)=1$ since a rescaling of $S$
corresponds to a trivial rescaling of the time coordinate.
Inserting a power series expansion into (16) one finds
$$\eqalignno{
W(r)&= 1-br^2+O(r^4)\;, \cr
N(r)&= 1-4b^2r^2+O(r^4)\;, \cr
S(r)&= 1+4b^2r^2+O(r^4)\;,&(17)}$$
where $b$ is an arbitrary parameter.

Similarly assuming a power series expansion in $1\over r$ at $r=\infty$
for asymptotically flat solutions, one finds
$\lim\limits_{r\to\infty}W(r)=\{\pm 1,0\}$.
It turns out that $W(\infty)=0$ cannot
occur for globally regular solutions, so we concentrate
on the remaining cases. One finds
$$\eqalignno{
 W(r)&= \pm \left(1-{c\over r}+O\left({1\over r^2}\right)\right)\;, \cr
 N(r)&= 1-{2M\over r}+O\left({1\over r^4}\right)\;, \cr
 S(r)&= S_\infty\left(1+O\left({1\over r^4}\right)\right)\;,&(18)}$$
where again $c, M$  and $S_\infty$ are arbitrary
parameters and have to be determined from numerical calculations.

It was found [9] that equations (16) admit
a discrete sequence of finite-energy solutions
$\{W_n, N_n, S_n \}$ which interpolate between
the asymptotic behaviours (17) for $r \to 0$
and (18) for $r \to \infty $.\medskip

\n {\bf 2.3. Dilatonic sphaleron}

As mentioned earlier, there are no static solutions in
the pure YM theory in $(3+1)$ dimensions. The reason is that
pure YM theory is repulsive. In order to have solutions
with finite energy one needs some extra field providing
attraction that compensates YM repulsion.
In the case of the electroweak sphaleron this attraction
is provided by a Higgs field.
It was realized [12], [13]
that the role of a binding force can
be provided by a dilaton field as well.

Introducing a dilaton field we naturally
obtain a EYMD theory with the action
$$S_{\rm EYMD}=
\frac{1}{4\pi}\int\left(-\, \frac{1}{4 G}R+
\frac{1}{2}(\partial\varphi)^2-
 {\e^{2\kappa\varphi}\over 4g^2}F^2\right)\sqrt{-g}\, d^4x \eqno (19)$$
where $\kappa$ and $g$ respectively denote the dilatonic and
gauge coupling constant and $G$ is Newton's constant.

This theory depends on a dimensionless parameter
$\gamma=\frac{\kappa}{g\sqrt{G}}$.
In the limit $\gamma\to 0$ one gets the EYM theory studied in
[9].
The value $\gamma=1$ corresponds to a model obtained from
heterotic string theory [19].
We found strong indications that the lowest-lying regular
solution for this value of  $\gamma$
may be obtained in closed form	[20], [21].

We will not discuss here the case of general $\gamma $
[20], [22],
but rather concentrate on the limiting case
$\gamma\to\infty$ where
one  obtains the YM-dilaton theory in
flat space  [12], [13].

In the flat case when gravity decouples
we get a YMD theory defined by the action
$$S_{\rm YMD}=
\frac{1}{4\pi}\int\left(
\frac{1}{2}(\partial\varphi)^2-
{\e^{2\varphi}\over 4g^2}F^2\right)d^4x. \eqno (20)$$

The corresponding reduced action is
$$S^{\rm red}_{\rm YMD}=-\int dr\left[{r^2\over 2}\varphi'^2+
\e^{2\varphi}\left(W'^2+{(W^2-1)^2\over 2r^2}\right)\right]\, \eqno (21)$$
with resulting field equations
$$\eqalignno{  W''&= {W(W^2-1)\over r^2}
	      -2\varphi' W'\;, \cr
 (r^2\varphi')'&=
 2\e^{2\varphi}\left(W'^2+{(1-W^2)^2\over2r^2}
 \right)\;.&(22)}$$
These equations are invariant under a shift
$\varphi\to\varphi+\varphi_0$ accompanied by a simultaneous rescaling
$r\to r\e^{\varphi_0}$. Hence
globally regular solutions can be normalized to $\varphi(\infty)=0$.

It was found [12], [13] that
equations (22) have a discrete sequence of finite energy
solutions $\{W_n, \varphi_n\}$, where $n=1, 2, 3,...$
labels the number of zeros of the gauge function $W_n(r)$.

The mass of this solution varies from $\approx 0.8$ for $n=1$
to $1.0$ for $n\to\infty$ in natural units $\kappa g$.

The asymptotic behavior of these solutions for $r\to 0$ is
$$\eqalignno{
W_{n}(r)&= 1-b_{n}r^2+O(r^4)\;, \cr
\varphi_{n}(r)&= \varphi_{n}(0)+
2\e^{2\varphi_{n}(0) }b_{n}^2r^2+O(r^4)\;,&(23)}$$
and for $r\to \infty$ is
$$\eqalignno{
W_{n}(r)&= (-1)^{n}\left(1-{c_{n}\over r}+O\left({1\over
r^2}\right)\right)\;, \cr
 \varphi_{n}(r) &= -{d_{n}\over r}+O\left({1\over r^4}\right)\;.&(24)}$$
The parameters $b_n,\varphi_n(0), c_n$ and $d_n$
have to be determined by numerical calculations.\vfill\eject

\n {\bf 2.4. Gravitating sphaleron}

A natural generalization of the YMH theory is
obtained taking gravity into account.
In this way one gets EYMH theory, which has
gravitating sphaleron [23], [24] solutions.

This theory contains two scales, gravitational and electroweak.
As a result one gets two kinds of excitation modes:
gravitational (analogous to the BMK mode) and electroweak.
The lowest solution is with one sphaleron node.
The next one, with  two nodes,
contains one sphaleron node and one gravitational node.\bigskip

\n {\bf 3. Properties and interpretation}

The main properties of the sphalerons are as follows:\medskip

\noindent
(i) they have finite energy

\noindent (ii) they have fractional topological charge

\noindent (iii) there are fermion zero modes in the background of these
solutions

\noindent (iv) they are saddle points of the action.\medskip

We call solutions of EYM and YMD  sphalerons since
they possess all the properties (i)--(iv).
More precisely,  solutions with odd $n$
are sphalerons,
while solutions with even $n$ correspond to  trivial
loops in configuration space.

(i) The mass of the sphaleron in the standard model of
electroweak interactions is of order of a few TeV.
The EYM sphalerons have typical masses of order  $1/g\sqrt{G}$.
If we assume that our model is part of the string theory,
or in other words if we are granted  the mass scale
parameter $M_{\rm Pl}$,
the mass of the solutions of the Einstein--Yang--Mills
theory is of the order of unity in Planckian  units.
In  the EYMD theory the mass of the solutions
decreases with increasing
dilatonic coupling constant $\gamma$ and
for large $\gamma$ goes to zero like
$M_{\rm EYMD}\sim \frac{M_{\rm Pl}}{\gamma ^2}$.

(ii) It was shown [2], [3]
how to assign a topological (baryon) number to the
sphaleron. It turns out to be ${1\over 2}$.
In the gauge we use here one can read this off from
the asymptotic behaviour of the gauge field.

(iii) It was found that there are fermion zero modes
in the background of the
electroweak [25], gravitational [26] and
dilatonic [27] sphalerons.
This is in perfect agreement with the picture of level-crossing phenomena [8].

(iv) The electroweak sphaleron has just one negative
mode [2], [3], [28].

Various aspects of the stability  of BMK solutions
have been analyzed [29], [30], [31], [32], [33], [34], [35].
It is natural [34]
to distinguish two different kinds of instabilities, which we call
`sphaleron' and `gravitational' instabilities.
Gravitational instabilities have no analogue for the flat-space sphaleron,
whereas instabilities of the former type have same nature
as for the YMH sphaleron.
It was found [29], [30] that the BMK solutions
are unstable and the numerical results for the few lowest solutions led to
the conclusion that the $n^{\rm th}$ Bartnik--McKinnon solution has exactly
$n$ unstable gravitational modes.  It was shown
analytically [31], [32]
that there exists at least one sphaleron-like unstable mode for
each member of the BMK family.

Numerical studies [34] led to
the claim that	the $n^{\rm th}$ BMK solution has exactly $n$
sphaleron-like instabilities, so that altogether the $n^{\rm th}$ BMK
solution has $2n$ unstable modes, $n$ of either type.  It is quite
remarkable that the conjecture about the number of  sphaleron-like
instabilities can be proven [35] in spite of the fact that the BMK
solutions are not known analytically.

Numerical studies indicates that the
same conjecture is true for dilatonic sphalerons, namely the $n^{\rm th}$
solution of the  YMD theory has exactly $2n$ unstable modes [27].\bigskip

\n {\bf 4. Comparison of different sphalerons}

It is an interesting question why such different theories as
YMH, EYM and YMD posses  similar solutions.
The ``explanation" lies in the presence of a YM field in all
of these cases.
Introducing a proper ``time" coordinate $\tau$ (different in each case)
one arrives at an equation of the following type:
$${d^2 W\over d\tau ^2} = -\, {d U\over d W}
+ \lambda{(\tau)}\dot{W} + h(\tau),\eqno (25)$$
where  $U(W)= - (W^2-1)^2/4$ is an inverted double
well potential, and the coefficients
$\lambda(\tau)$ and $h(\tau)$ for each	theory
are shown in the Table.

$$\vbox{\offinterlineskip \halign{\strut \vrule \ \ #\hfil \ \ &\vrule \
\ $#$\hfil \ \ &\vrule \ \ $#$\hfil \ \ \vrule\cr
\noalign{\hrule}
Theory	& \lambda(\tau) &  h(\tau)    \cr
\noalign{\hrule}
YM	&  1		& 0	    \cr
YMH	&  1		& {e^{2\tau} H^2\over 4}(1+W)  \cr
EYM & K-{\dot{K}\over K}-{2{\dot{W}}^2 \over K r} & 0 \cr
YMD	& 1-2\dot{\varphi} & 0	      \cr
\noalign{\hrule}}}$$
\centerline{{\bf Table.} Coefficients in equation (25) for different
theories. $K \equiv \sqrt{N}$.}\medskip
Without  the last term $h(\tau)$ the equation (25)
has a simple mechanical analogue, the motion
of a ``particle" in the potential $U(W)$ under the influence of
friction, with the coefficient	$\lambda(\tau)$.
The last term, which is
present only in the YMH case, plays the role of a ``time dependent"
potential.

In order to have a finite-energy field-theoretical solution
the ``particle" in our mechanical analogy should start at $W=1$
for $\tau\to -\infty$ with $\dot{W}=0$ and end up at $W=\pm 1$
for $\tau =+\infty$.

It is obvious that there are no static solutions in the pure
YM theory, because the constant friction term
prevents the ``particle" from stopping at the top, $W=-1$.

In the YMH case the friction term is the same but the potential is
deformed with increasing $\tau$.

In EYM and YMD cases the friction coefficient depends
on the derivatives of other fields.
Thus one can have positive as well as negative friction.
This allows for  excited solutions in the cases with
gravity and dilaton, corresponding to oscillations of the
``particle"\negthinspace.

Although all the solutions we discuss are similar, there is an
important difference.
In contrast to the electroweak sphaleron
the gravitational and dilatonic sphalerons have an even number
of negative modes  [34], [27].
((This fact should be considered in  view of the recent paper of
Rubakov and Shvedov [36].))\bigskip

\n {\bf 5. Concluding remarks}

We have shown that there are  saddle point
solutions of similar nature  in YMH, EYM and YMD theories.
We emphasized the similarities and differences between these solutions.

In addition to the globally regular solutions considered above
there are also black hole solutions [37], [20], [23]
of corresponding theories.
They may be considered as black holes sitting inside sphalerons.
These  solutions are of interest as  counterexamples for
a ``no-hair" conjecture.

The  ``zoo"  of solutions described is an interesting problem of
mathematical physics.
The sphaleron is important in  electroweak baryogenesis.
The role and importance of the gravitational and dilatonic
analogues is not yet clear.\bigskip

\n {\bf 6. Acknowledgments}

I wish to thank  Gabor Kunstatter and Tom Osborn for the invitation
to the Heat Kernel Technique and Quantum Gravity conference
and warm hospitality.
I am  grateful to Dieter Maison and Norbert Straumann
for many useful discussions.
This work was supported in part by the Tomalla Foundation.
\bigskip
\vfill\eject

\n {\bf References}\medskip

\frenchspacing

\item{[1]} R. Jackiw and C. Rebbi,
{\it Phys. Rev. Lett.} {\bf 37} (1976) 172;
\item{}C.G. Callan, R.F. Dashen, and D.J. Gross,
{\it Phys. Lett.} {\bf 63B} (1976) 334.

\item{[2]} N. Manton,
{\it Phys. Rev.} {\bf D28} (1983) 2019.

\item{[3]} F.R. Klinkhamer and N.S. Manton ,
{\it Phys. Rev.} {\bf D30} (1984) 2212.

\item{[4]} G. 't Hooft,
{\it Phys. Rev. Lett.} {\bf 37} (1976) 8;
\item{} {\it Phys. Rev.} {\bf D14} (1976) 3432.

\item{[5]} S.L. Adler,
{\it Phys. Rev.} {\bf 177} (1969) 2426;
\item{} J.S. Bell and R. Jackiw,
{\it Nuovo Chimento} {\bf 60A} (1969) 47.

\item{[6]}N.H. Christ,
{\it Phys. Rev.} {\bf D21} (1980) 1591.

\item{[7]} G. Lavrelashvili,
{\it Theor. Math. Phys.} {\bf 73}  (1987) 1191.

\item{[8]} J. Kiskis,
{\it Phys. Rev.} {\bf D18} (1978) 3690.

\item{[9]} R. Bartnik and J. McKinnon,
 {\it Phys.\ Rev.\ Lett.\/} {\bf 61} (1988) 141.

\item{[10]}  M.S. Volkov and D.V. Gal'tsov, {\it  Phys. Lett.}
{\bf B273} (1991) 255.

\item{[11]} D. Sudarsky and R.M. Wald, {\it Phys. Rev.}
{\bf D46} (1992) 1453.

\item{[12]} G. Lavrelashvili and D. Maison,
{\it Phys. Lett.} {\bf B295} (1992) 67.

\item{[13]} P. Bizon,
{\it Phys. Rev.} {\bf D47} (1993) 1656.

\item{[14]} R.F. Dashen, B. Hasslacher, and A. Neveu,
{\it Phys. Rev.} {\bf D10} (1974) 4138.

\item{[15]} J. Boguta,
{\it Phys. Rev. Lett.} {\bf 50} (1983) 148.

\item{[16]} S. Coleman, in {\it New Phenomena in Subnuclear Physics,}
 A. Zichichi ed., Plenum, New York, 1975;
\item{} S. Deser, {\it Phys. Lett.} {\bf B64} (1976) 463.

\item{[17]} E. Witten,
{\it Phys. Rev. Lett. }
{\bf 38} (1977) 121;
\item{} P. Forg\'acs and  N.S. Manton,
{\it Commun. Math. Phys. } {\bf 72} (1980) 15.

\item{[18]} S. Deser, {\it Class. Quantum Grav. } {\bf 1} (1984) L1.

\item{[19]} M.B. Green, J.H. Schwarz and E. Witten,
{\it Superstring Theory\/} (Cambridge U.P., Cambridge, 1987).

\item{[20]}  G. Lavrelashvili and D. Maison,
 {\it Nucl.\ Phys.\/} {\bf B410}  (1993) 407.

\item{[21]}
G. Lavrelashvili,
{\it Black holes and sphalerons in low energy string theory},
 hep-th/9410183.

\item{[22]}
 P. Bizon,
 {\it Acta Physica Polonica\/} {\bf B24} (1993) 1209;
\item{} E.E. Donets and D.V. Gal'tsov,
 {\it Phys.\ Lett.\/} {\bf B302} (1993) 411;
\item{}  T. Torii and Kei-ichi Maeda,
 {\it Phys.\ Rev.\/} {\bf D48} (1993) 1643;
\item{} O'Neill, {\it Phys.\ Rev.\/} {\bf D50} (1994) 865.

\item{[23]}
 B.R. Greene, S.D. Mathur and C.M. O'Neill,
 {\it Phys. Rev.}
 {\bf D47} (1994) 2242.

\item{[24]}
P. Breitenlohner, G. Lavrelashvili and D. Maison, (1994),
unpublished.

\item{[25]}
A. Ringwald,
{\it Phys. Lett.} {\bf B213} (1988) 61;
\item{} J. Kunz and Y. Brihaye,
{\it Phys. Lett.} {\bf B304} (1993) 141.

\item{[26]}
G.W. Gibbons and A.R. Steif,
{\it Phys. Lett.} {\bf B314} (1993) 13;
\item{} M.S. Volkov,
{\it Phys. Lett.} {\bf B334} (1994) 40.

\item{[27]}
G. Lavrelashvili,
{\it Mod. Phys. Lett.} {\bf A9} (1994) 3731.

\item{[28]}
J. Burzlaff, {\it  Nucl. Phys.}
{\bf B233} (1984) 262;
\item{} L. G. Yaffe, {\it Phys. Rev.}
{\bf D40} (1989) 3463.


\item{[29]}
N. Straumann and Z.H. Zhou, {\it Phys. Lett.}
{\bf B243} (1990) 33.

\item{[30]}
Z.H. Zhou, {\it Helv. Phys. Acta}
{\bf 65} (1992) 767.

\item{[31]}
O. Brodbeck and N. Straumann,
{\it Phys. Lett.} {\bf B324} (1994) 309.

\item{[32]}
M.S. Volkov and D.V. Gal'tsov, {\it Odd-parity negative modes of
Einstein-Yang-Mills black holes and sphalerons,}
Preprint ZU-TH 27/94.

\item{[33]}
T. Torri, K. Maeda, T. Tachizawa,
{\it Phys. Rev.} {\bf D51} (1995) 1510.

\item{[34]}
G. Lavrelashvili and D. Maison,
{\it Phys. Lett.} {\bf B343} (1995) 214.

\item{[35]}
M.S. Volkov, O. Brodbeck, G. Lavrelashvili and N. Straumann,
{\it Phys. Lett.} {\bf B349} (1995) 438.

\item{[36]}
V.A. Rubakov and O.Yu. Shvedov,
{\it Nucl. Phys.} {\bf B434} (1995) 245.

\item{[37]}
 M.S. Volkov and D.V. Gal'tsov,
{\it  JETP\ Lett.\/} {\bf 50} (1990) 346;
\item{} H.P. K\"unzle and A.K.M. Masood-ul-Alam,
{\it J.\ Math.\ Phys.\/} {\bf 31} (1990) 928;
\item{} P. Bizon,
{\it Phys.\ Rev.\ Lett.\/ } {\bf 61} (1990) 2844.
\bye